# A Proof Of Kochen - Specker Theorem of Quantum Mechanics Using a Quantum Like Algebraic Formulation

## Elio Conte

*Department of Pharmacology and Human Physiology and Tires, Center for InnovativeTechnologies for Signal Detection and Processing, University of Bari, Italy;*
*School of Advanced International Studies on Theoretical and nonLinear Methodologies, Bari, Italy*

**Abstract:** using a quantum like algebraic formulation we give proof of Kochen-Specker theorem. We introduce new criteria in order to account for the contextual nature of measurements in quantum mechanics.

## 1.Introduction

It is well known that the definition of Quantum Like may assume different interpretations according to various authors. In this paper we will intend by this term a rough scheme of quantum mechanics, a bare bone skeleton of the fundamental theory, as it may be introduced by using the Clifford algebra. It was discussed in a number of previous papers [1], and we will no more consider it in detail in the present work. The reader is send back to such papers in order to deepen this argument. We will consider here only some algebraic quantum like notations just to enable the reader to follow our elaboration.

## 2. A rough Quantum Like Scheme of Quantum Mechanics

Let us introduce two basic algebraic axioms:
1) $e_i$, $i = 1,2,3$ are abstract – symbolic-algebraic elements whose square gives 1,
$$e_i^2 = 1. \tag{1}$$
2) Such algebraic elements are anti commutative, that is to say:
$$e_i e_j = -e_j e_i, \qquad i = 1,2,3; j = 1,2,3; i \neq j. \tag{2}$$
Only these two algebraic axioms, (1) and (2), are required [1] in order to realize a rough quantum mechanical scheme. In particular, we will not use the concept of wave function that is at the basis of quantum mechanical formalism.

Let us comment on the axioms (1) and (2). Here the problem is only of interpretation. We call (1) the axiom of the potentiality. In fact, owing to the axioms (1) and (2), the $e_i$ cannot represent any number in some given field. However, owing to the axiom (1), each $e_i$ in (1) has the potentiality that we could attribute to it a numerical value, that is or +1 or -1.

Usual quantum mechanics also introduces a net distinction between potentiality and actualization from its starting and, in particular, through the superposition principle of quantum states. In our quantum mechanical rough scheme we have the potential superposition of ($\pm 1$) values for each of the given elements specified in (1).

A theorem, discussed in [1], proves the existence of the algebra given by (1) and (2). The two given axioms, (1) and (2), are sufficient to realize an algebraic structure that we call the Algebra A.
In particular, such proof shows that we have
$$e_1 e_2 = -e_2 e_1 = i e_3, \; e_2 e_3 = -e_3 e_2 = i e_1, e_3 e_1 = -e_1 e_3 = i e_2 \tag{3}$$
that is based on a cyclic permutation $(i\;j\;k)$ of $(1\;2\;3)$. All the basic features of this algebra pass through such fundamental requirement: a cyclic permutation $(i\;j\;k)$ of $(1\;2\;3)$ is strongly required. This is the algebra A and we send back to ref.[1] for all its features.

In our interpretation it represents the algebra of the potentiality. The sense of this notion is that we never may give a direct numerical value to the algebraic elements or members of this algebra but, however, such mathematical structure preserves at any time the potentiality that some numerical value could be attributed to its basic elements. The theorem 2, discussed in [1], proves that we may also realize a passage from potentiality to actualization. We are in the condition to attribute a definite numerical value to the basic algebraic elements identified in (1) and (2), and in this case we have a passage from the Algebra A to a subalgebra B. This subalgebra B is characterized by the fact that to one and only to one element, $e_i$ ($i = 1,2,3$), we may attribute a definite numerical value, say o +1 or -1 while at the same time the other elements remain completely undetermined under the profile of their numerical value. A new algebraic set is obtained in the algebra B. Considering, as example, to attribute the +1 value to $e_3$, we have subalgebra B characterized in the following manner

*subalgebra B :*
$e_1^2 = 1$, $e_2^2 = 1$, $i_B^2 = -1$;
$e_1 e_2 = i_B$, $e_2 e_1 = -i_B$, $e_2 i_B = -e_1$, $i_B e_2 = e_1$, $e_1 i_B = e_2$, $i_B e_1 = -e_2$ ; $e_3 \to 1$ (4)

The theorem 2 describes the passage from potentiality to actualization.
In quantum mechanics we have the passage from potentiality to actualization by the so called mechanism of wave function reduction that, however, it is only admitted but never derived in the framework of such theory.
Let us see now some other features in such algebraic scheme. In algebra A, considering that each element $e_i$ has the potentiality to assume the numerical value of +1 or of -1, we are implicitly admitting that the $e_i$ are characterized from an intrinsic indetermination. This is to say that at the stage of their potentiality, there exists an ontological probability $p_{+1}$ for each $e_i$ to assume the value +1 and an ontological probability $p_{-1}$ to assume the value -1, given in the following manner:

$$p_{+1} = \frac{1}{2} + <e_i> \quad \text{and} \quad p_{-1} = \frac{1}{2} - <e_i> \quad (5)$$

being $<e_i>$ the mean value for the basic element $e_i$.
In this manner we delineate a rough quantum scheme through an algebraic structure. Starting with the general validity of quantum mechanics, the axioms (1) and (2) could represent the mathematical counterpart of some basic feature of our reality, that one of potentiality.
Let us add still some other mathematical notation. The basic elements $e_i$ remain substantially abstract and indefinite in the algebraic structures A and B. We may take advantage of a practical representation of such basic abstract elements $e_i$ by an isomorphism.
The isomorphism of such basic elements with matrices at $n = 2$ is in fact well known. They are given in the following manner:

$$e_1 = \begin{pmatrix} 0 & 1 \\ 1 & 0 \end{pmatrix}, \quad e_2 = \begin{pmatrix} 0 & -i \\ i & 0 \end{pmatrix} \quad \text{ed} \quad e_3 = \begin{pmatrix} 1 & 0 \\ 0 & -1 \end{pmatrix} \quad (6)$$

By using direct product of matrices we may realize sets of basic elements at different orders n. As example, at n = 4, we have that

$$E_{01} = 1 \otimes e_1 = \begin{pmatrix} 1 & 0 \\ 0 & 1 \end{pmatrix} \otimes \begin{pmatrix} 0 & 1 \\ 1 & 0 \end{pmatrix} = \begin{pmatrix} 0 & 1 & 0 & 0 \\ 1 & 0 & 0 & 0 \\ 0 & 0 & 0 & 1 \\ 0 & 0 & 1 & 0 \end{pmatrix} . E_{10} = e_1 \otimes 1 = \begin{pmatrix} 0 & 1 \\ 1 & 0 \end{pmatrix} \otimes \begin{pmatrix} 1 & 0 \\ 0 & 1 \end{pmatrix} = \begin{pmatrix} 0 & 0 & 1 & 0 \\ 0 & 0 & 0 & 1 \\ 1 & 0 & 0 & 0 \\ 0 & 1 & 0 & 0 \end{pmatrix}$$

$$E_{02} = \begin{pmatrix} 0 & -i & 0 & 0 \\ i & 0 & 0 & 0 \\ 0 & 0 & 0 & -i \\ 0 & 0 & i & 0 \end{pmatrix}, \qquad E_{03} = \begin{pmatrix} 1 & 0 & 0 & 0 \\ 0 & -1 & 0 & 0 \\ 0 & 0 & 1 & 0 \\ 0 & 0 & 0 & -1 \end{pmatrix},$$

$$E_{20} = \begin{pmatrix} 0 & 0 & -i & 0 \\ 0 & 0 & 0 & -i \\ i & 0 & 0 & 0 \\ 0 & i & 0 & 0 \end{pmatrix}, \qquad E_{30} = \begin{pmatrix} 1 & 0 & 0 & 0 \\ 0 & 1 & 0 & 0 \\ 0 & 0 & -1 & 0 \\ 0 & 0 & 0 & -1 \end{pmatrix} \tag{7}$$

In correspondence of the (1) and the (2) we have now that
$$E_{01}^2 = E_{02}^2 = E_{03}^2 = 1 \quad \text{and} \quad E_{01}E_{02} = iE_{03}, \; E_{02}E_{03} = iE_{01}, \; E_{03}E_{01} = iE_{02}$$
and equivalent relations for $E_{j0}, \; j = 1,2,3$. \hfill (8)

In the same manner we may proceed at different orders n.
Let us observe that some different basic sets of the algebra A at order n=4 may be considered.
They are given in the following manner:

*Order n=4. The following twenty algebraic sets are given:*
(1) – $(E_{01}, E_{02}, E_{03})$, (2) – $(E_{10}, E_{20}, E_{30})$,
(3) – $(E_{01}, E_{12}, E_{13})$, (4) – $(E_{01}, E_{22}, E_{23})$, (5) – $(E_{01}, E_{32}, E_{33})$, (6) – $(E_{11}, E_{02}, E_{13})$, (7) – $(E_{21}, E_{02}, E_{23})$,
(8) – $(E_{31}, E_{02}, E_{33})$, (9) – $(E_{11}, E_{12}, E_{03})$,
(10) – $(E_{21}, E_{22}, E_{03})$, (11) – $(E_{31}, E_{32}, E_{03})$, (12) – $(E_{10}, E_{23}, E_{33})$, (13) – $(E_{10}, E_{22}, E_{32})$, (14) – $(E_{10}, E_{21}, E_{31})$,
(15) – $(E_{12}, E_{20}, E_{32},)$, (16) – $(E_{11}, E_{20}, E_{31})$, (17) $(E_{13}, E_{20}, E_{33})$
(18) – $(E_{13}, E_{23}, E_{30})$, (19) – $(E_{12}, E_{22}, E_{30})$, (20) – $(E_{11}, E_{21}, E_{30})$. \hfill (9)

These are the only possible basic sets at the order (n=4) with $E_{ij} = E_{i0}E_{0j}$.

It is very important to observe that each basic set is regulated by only one cyclic permutation $(i \; j \; k)$ of $(1,2,3)$ and $(j \; j \; 0)$ or $(j \; 0 \; j)$ or $(0 \; j \; j)$ with $(j = 1,2,3)$.

*These are the basic rules of our algebra: Each basic set has its characterizing basic permutation and its characterizing $(j \; j \; 0)$ with $(j = 1,2,3)$.*

These are the logic and algebraic consenquences to have admitted the two starting postulates in (1) and (2), where, in particular, the (2) relates anti commutativity of such elements. If, for example, we consider the (20) in (9), we have that
$$E_{11}^2 = E_{21}^2 = E_{30}^2 = 1 \tag{10}$$
and
$$E_{11}E_{21} = iE_{30}, \; E_{21}E_{30} = iE_{11}, \; E_{30}E_{11} = iE_{21} \quad \text{and}$$
$$E_{11}E_{21} = -E_{21}E_{11}, \; E_{21}E_{30} = -E_{30}E_{21}, \; E_{11}E_{21} = -E_{21}E_{11} \tag{11}$$

In this case we have the cyclic permutation of (1 2 3) as required and $(j \; j \; 0)$ as basic rule.

Non commutativity from one hand and potentiality as expressed by (1) are thus the fundations of our algebraic structure. We may say that each set, from (1) to (20) in (9), identifies an abstract space that would be the space of quantum like events in which potentiality, expressed by the (1) and the (2), or equivalently, by the (8), operates. We have the algebraic PRESPACE of potentialities actualized after in the ordinary space of our experience.

There is still another feature that appears to be of importance in the (9). Let us give up for a moment the pure algebraic language and let us connect it to the sphere of the experience. Let us admit to have a system of ½ spin particles. $(E_{01}, E_{02}, E_{03})$ will be the spin for particle (a) and

$(E_{10}, E_{20}, E_{30})$ will be the spin for particle (b). Let us admit that in such quantum system we have to measure $E_{03}$. Looking at the (9), one deduces that, if we have to respect the basic algebraic features that preside over to our algebra, we are free to consider only some prefixed algebraic sets that, in the case of $E_{03}$, are

(1) – $(E_{01}, E_{02}, E_{03})$, (9) – $(E_{11}, E_{12}, E_{03})$, (10)-$(E_{21}, E_{22}, E_{03})$, (11)-$(E_{31}, E_{32}, E_{03})$ (12)

Let us remember that we are discussing a pure scheme of potentiality. Each algebraic set of the (12) indicates the group of anticommuting spin components that we could not measure simultaneously with $E_{03}$.

In the first case, the (1) indicates that we could not measure $E_{01}$ or $E_{02}$, measuring $E_{03}$. In other terms, we have a scheme of potentiality such that an experimental arrangement could actualize or $E_{03}$ or $E_{02}$ or $E_{01}$.

In the second case, the (9) in (12) indicates that we could actualize or $E_{03}$ or $E_{10}$ and $E_{02}$ or $E_{10}$ and $E_{01}$. In the third case we could actualize or $E_{03}$ or $E_{20}$ and $E_{01}$ or $E_{20}$ and $E_{02}$. Finally, in the last case we could actualize $E_{03}$ or $E_{30}$ or $E_{01}$ or $E_{30}$ and $E_{02}$.

We have reached rather obvious conclusions. Also the usual quantum mechanics indicates each time what we may measure or not by using the language of operators in Hilbert spaces, and considering each time pairs of commuting or non commuting operators. The only difference is that in our algebraic framework we have the direct indication of each algebraic set, and the basic cyclic permutations of the basic elements that are involved in each algebraic set. In this manner we have a clear and rigorous indication of the fields of potentiality that are involved during the measurement of $E_{03}$.

The same thing happens when admitting to measure $E_{30}$. In this case the basic algebraic sets that would be involved, will be the (2), the (12), the (13), and the (14) of the (9).

Let us admit now, instead, that we intend to measure $E_{33}$, where $E_{03}E_{30} = E_{33}$. We have the following sets: (5), (8), (12), (17) of the (9). If, for example, we are in the (5), we may actualize or $E_{03}$ and $E_{30}$ or $E_{01}$ or $E_{30}$ and $E_{02}$.

Now let us go back to the theorem we showed in [1] on transition potentiality-actualization and restart considering the basic elements of our algebra at order n=2 as given in (3). Our theorem shows that, if we start with the algebra A that is given in (1) and (2), if we actualize as example $e_3$, we have a passage from algebra A to the subalgebra B where the new basic elements are given in the following manner

*subalgebra B:*
$e_1^2 = 1$, $e_2^2 = 1$, $i_B^2 = -1$;
$e_1 e_2 = i_B$, $e_2 e_1 = -i_B$, $e_2 i_B = -e_1$, $i_B e_2 = e_1$, $e_1 i_B = e_2$, $i_B e_1 = -e_2$ ; $e_3 \to 1$ (13)

with equivalent relations for $e_3 \to -1$.

## 3. The Proof of Kochen-Specker Theorem

The substantial difference between algebra A corresponding to the regime of potentiality to algebra B corresponding to the regime of actualization of one of the three basic elements (in this case it is considered $e_3$) is in the role of the unit $i$ that in the case of algebra A is given by $e_1 e_2 e_3 = i$ while, when going to actualization, it becomes a basic elements of the new algebra, $i_B$, as given in (11).

Let us explore what happens in the case of the quantum two ½ spin particles previously considered. In the case of $E_{03}$ actualization, we have the (12). For each algebraic set, we may calculate the corresponding $i_B$ basic element and we obtain that in all the four algebraic sets shown in (12), it is given in the following manner

$$i_B = E_{01} E_{02} \tag{14}$$

Note that we have the same $i_B$ for each of the basic algebraic sets given in (12).

Let us admit now that in our quantum system, we give actualization to $E_{30}$ and not to $E_{03}$. As previously discussed, we will have the basic sets named (2), (12), (13), (14) of the (9). As expected, in this case $i_B$ will be again the same for all the four algebraic sets, and it will be given in the following manner

$$i_B = E_{10} E_{20} \tag{15}$$

The (14) and the (15) are in perfect accord with the (13) as expected in the case of actualization of only one spin component $E_{03}$ considering particle (a) or $E_{30}$ considering particle (b).

Let us consider now the case of a simultaneous actualization of $E_{03}$ and $E_{30}$. We know that the basic algebraic sets are given this time in (5), (8), (12) and (17) of the (9).

Note that in this case we will have two different $i_B$, and precisely :

For the basic sets $(E_{01}, E_{32}, E_{33})$ and $(E_{31}, E_{02}, E_{33})$ it will be that

$$i_B = E_{30} E_{01} E_{02} \tag{16}$$

Instead, for the basic sets $(E_{10}, E_{23}, E_{33})$ and $(E_{13}, E_{20}, E_{33})$ it will be that

$$i_B = E_{03} E_{10} E_{20} \tag{17}$$

We may now give the results in the following manner:

$$i_{B(E_{33})} \neq i_{B(E_{03})} \neq i_{B(E_{30})} \tag{18}$$

and

$$i_{B(E_{33})} = i_{B(E_{03})} E_{30}$$
$$i_{B(E_{33})} = i_{B(E_{30})} E_{03} \tag{19}$$

Therefore, we conclude that, given

$$E_{33} = E_{30} E_{03} \tag{20}$$

in some way we may consider that they are equivalent the actualizations of $E_{03}$, $E_{30}$, and $E_{33}$. Under their algebraic profile the (19) both value simultaneously, they are one the counterpart of the other and are strongly linked to (14 ) and (15).

We outline here that the corresponding question for usual quantum mechanics is posed for given operators $A$, $B$, and $[A, B] = 0$ in the actualization of their corresponding observables $A, B, AB$. Note that the (14) and the (15), and, consequently, the (19) indicate two different schemes of possible intrinsic indetermination.

Consider now $E_{30}, E_{03}$, and $E_{02}$. We know that $E_{30}$ and $E_{03}$ are commutative. $E_{30}$ and $E_{02}$ are commutative but $E_{03}$ and $E_{02}$ are not.

When we actualize $E_{30}$ we have the (15) that is

$$i_B = E_{10} E_{20} \tag{21}$$

When we actualize $E_{33} = E_{30} E_{03}$, we have the (16) and the (17) that is

$$i_B = E_{30} E_{01} E_{02} \tag{22}$$

and

$$i_B = E_{03} E_{10} E_{20} \tag{23}$$

When we actualize $E_{32} = E_{30} E_{02}$, we have

$$i_B = E_{10} E_{20} E_{02} \tag{24}$$

and

$$i_B = -E_{30}E_{01}E_{03} \tag{25}$$

By comparison of the (24) and (25) with the (22) and (23) and with the (21), we conclude that we cannot reconcile the (24) and the (25) with the (22) and the (23) and with the (21). Therefore, we may conclude that it is not the same thing to actualize $E_{30}$ or $E_{30}$ and $E_{03}$ or $E_{30}$ *and* $E_{02}$.

According to various authors, and in particular to A. Peres [2], there is a quantum analog in quantum mechanics as it is usually called the necessity of contextual measurements in quantum mechanics. Also its link with the Kochen-Specker theorem is well known, and thus it will not be further outlined here. We retain that it may be of some relevance the fact that we are giving here a proff and thus a rigorous derivation, an elaboration and a physical significance of such quantum results on the basis of an algebraic framework that of course is based only on two axioms as given by the (1) and (2) and some following theorems [3].

Let us continue giving the proof on the basis of an equivalent algebraic formulation.

Consider this time $E_{03}$, $E_{30}$, $E_{01}$, and $E_{10}$. Again we have that $E_{03}E_{30} = E_{30}E_{03}, E_{30}E_{01} = E_{01}E_{30}, E_{03}E_{10} = E_{10}E_{03}, E_{10}E_{01} = E_{01}E_{10}$. The other remaining pairs of algebraic elements are obviously not commutative.

Let us decide to actualize $E_{13} = E_{10}E_{03}$. Looking at the (9) and operating on the basis of our previous theorem [1], we have that it must be

$$i_B = E_{10}E_{01}E_{02} \tag{26}$$

and

$$i_B = E_{03}E_{20}E_{30} \tag{27}$$

Consider to actualize $E_{31} = E_{30}E_{01}$. Following the same procedure we have that it must be

$$i_B = E_{10}E_{20}E_{01} \tag{28}$$

and

$$i_B = E_{02}E_{03}E_{30} \tag{29}$$

Let us actualize now $E_{11} = E_{10}E_{01}$. It will be that

$$i_B = E_{10}E_{02}E_{03} \tag{30}$$

and

$$i_B = E_{01}E_{20}E_{30} \tag{31}$$

Finally, let us actualize $E_{33} = E_{30}E_{03}$. We know that it will be

$$i_B = E_{30}E_{01}E_{02} \tag{32}$$

and

$$i_B = E_{03}E_{10}E_{20} \tag{33}$$

Let us observe the resemblance of the (30) relating actualization of $E_{11}$, with the (29) relating $E_{31}$ but unfortunately $E_{10}$ and $E_{30}$ are involved that on the other hand are not commutative. The same thing happens for the (28) of $E_{31}$ and the (33) of $E_{33}$. The same question is for the (31) of $E_{11}$ with the (27) of $E_{13}$. Finally, the same problem arises for the (26) of $E_{13}$ and the (32) of $E_{33}$. Owing to such non commutativities, also starting with prefixed numerical values, given for actualization of commutative $E_{03}$ and $E_{10}$, and $E_{30}$ and $E_{01}$ or $E_{03}$ and $E_{30}$ and $E_{10}$ and $E_{01}$, we cannot arrive to attribute final definite values to the remaining basic elements.

Let us give proof of the theorem in detail.

Let us admit that we would intend to introduce two new basic algebraic sets, starting with $E_{11}$ and $E_{33}$ from one hand and with $E_{13}$ and $E_{31}$ from the other hand. We could think to admit the following two new basic sets

(1)- ($E_{11}, E_{22}, E_{33}$)   and   (2)-($E_{13}, E_{22}, E_{31}$) \qquad (34)

Note that all the basic algebraic sets that we introduced previously in (9), all they had only one
cyclic permutation $(i, j, k)$ of $(1,2,3)$ (35)
and
a basic rule $(0j\ j)$ or $(j\ 0\ j)$ or $(j\ j\ 0)$. (36)

All the algebraic sets that we introduced in (9) had such two basic requirements that are indispensable in our algebraic formulation and in the corresponding quantum like formulation.

Let us observe now the sets introduced in (34).

Here we are changing radically our algebraic rules in the sense that this time, instead of the (35) and the (36), we are considering two cyclic permutations: in (1) of the (34) we are considering a double permutation $(i\ j\ k)$ of $(1\ 2\ 3)$ in $E_{11}$, $E_{22}$, and $E_{33}$ while in (2) of the (34) we are considering two inverted permutations $(i\ j\ k)$ and $(k\ j\ i)$ of $(1, 2, 3)$ in $E_{13}$, $E_{22}$, $E_{31}$. We are completely changing our algebraic approach to the quantum like representation.

In addition, it seems that we have an advantage respect to traditional quantum mechanics, in the case of our algebraic quantum like formulation because we are supported here from algebraic criteria that instead are missing in the traditional quantum language of operators and observables. In other terms, before to admit, in case, the two new basis sets given in (34), we should ask preliminarily if they should be or should not be admissible on the basis of the two axioms given in (1) and (2).

If we should introduce the new two basic sets given in (34), following the criteria used in the previous proof in [1], we must ascertain that they verify all the features of our algebra. The reader is sent back to ref. [1] to inspect the technique of the proof. In the case of the (1) in (34) we will have:

$$E_{11}E_{22} = E_{10}E_{01}E_{20}E_{02} = (\gamma_1 E_{10} + \gamma_2 E_{20} + \gamma_3 E_{30})(\omega_1 E_{01} + \omega_2 E_{02} + \omega_3 E_{03})$$ (37)

Following step by step the technique exposed in ref.[1], one establishes that the (37) holds for

$$\gamma_1 = \gamma_2 = 0\ ; \gamma_3 = i\quad \text{and}\quad \omega_1 = \omega_2 = 0\ ;\omega_3 = i$$ (38)

For

$$E_{22}E_{33} = E_{20}E_{02}E_{30}E_{03} = (\alpha_1 E_{10} + \alpha_2 E_{20} + \alpha_3 E_{30})(\beta_1 E_{01} + \beta_2 E_{02} + \beta_3 E_{03})$$ (39)

one finds that it holds for

$$\alpha_1 = i\ ; \alpha_2 = \alpha_3 = 0\quad \text{and}\quad \beta_1 = i\ ; \beta_2 = \beta_3 = 0$$ (40)

Finally, for

$$E_{33}E_{11} = E_{30}E_{03}E_{10}E_{01} = (\vartheta_1 E_{10} + \vartheta_2 E_{20} + \vartheta_3 E_{30})(\chi_1 E_{01} + \chi_2 E_{02} + \chi_3 E_{03})$$ (41)

it holds for

$$\vartheta_1 = \vartheta_3 = 0\ ; \vartheta_2 = i\quad \text{and}\quad \chi_1 = \chi_3 = 0\ ; \chi_2 = i$$ (42)

In brief, we obtain that

$$E_{11}E_{22} = E_{22}E_{11} = -E_{33}\ ;\ E_{22}E_{33} = E_{33}E_{22} = -E_{11}\ ;\ E_{33}E_{11} = E_{11}E_{33} = -E_{22}$$ (43)

We may now summarize the obtained results. We started with $E_{11}$ and $E_{33}$. We observed that, in addition to the basic sets that we introduced in (7) and all observing the rules given in (35) and in (36), we could still admit to consider other two basic sets:

(1)- $(E_{11}, E_{22}, E_{33})$ and (2)-$(E_{13}, E_{22}, E_{31})$ (44)

previously written in (34).

We verified that the (1) of (44) is not admissible in our algebraic framework first of all because the basic elements $E_{11}$, $E_{22}$, and $E_{33}$ result to be commutative instead to be anticommutative and, in addition, the basic rule for multiplication of such basic elements, results to be violated since their product, see the (43), is no more of the kind given in (11) but a minus sign appears instead of imaginary unity, $i$. In brief, if we use the (1) of the (44) we are out of our quantum like algebraic scheme, and, in particular, we violate the basic rule of non cummativity that is at the foundation of such algebra. At the same time we retain that it is out of any possible quantum mechanical scheme.

Using it we should come into some algebraic contradiction in our quantum like algebraic framework and in some conceptual difficulty using it in traditional quantum mechanics.

Let us examine now the (2) of the (44). Using the same procedure we finally conclude that

$$E_{13}E_{22} = E_{22}E_{13} = E_{31} \ ; \ E_{22}E_{31} = E_{31}E_{22} = E_{13} \ ; \ E_{31}E_{13} = E_{13}E_{31} = E_{22} \tag{45}$$

Also this quantum like algebraic set is not admissible. It results commutative and again it violates the basic rules of multiplication of the elements as previously seen for the (43). Using this set certainly we are out of our quantum like algebraic formulation, and we are also out from usual quantum mechanical scheme. Using it, we should come into some quantum like algebraic contradiction.

In fact, such contradiction is rather evident. Consider the basic set (1) given in (44). In order to actualise some algebraic element we should actualize $E_{10}, E_{01}, E_{30}$, and $E_{03}$. We could not actualize all such elements at the same time owing to the non commutativity. So, we could choice to actualise one time $E_{10}$ and $E_{01}$ and in another time $E_{30}$ and $E_{03}$. Considering the (2) in (44), we should actualise one time $E_{10}$ and $E_{03}$, and in another time $E_{30}$ and $E_{01}$. In any case the knowledge of $E_{10}, E_{01}, E_{30}$, and $E_{03}$ is required. With such prefixed numerical values attributed to such elements, we should be able to obtain from one hand

$$E_{22} = -E_{33}E_{11} = -E_{30}E_{03}E_{10}E_{01} \tag{46}$$

and from the other hand

$$E_{22} = E_{31}E_{13} = E_{30}E_{01}E_{10}E_{03} \tag{47}$$

and this is a manifest algebraic contradiction. We cannot actualize, that is to say, we cannot attribute such numerical values in accord with the modalities that we have forced to consider in such algebraic scheme. In particular, we cannot claim to have coherent results using one time values for $E_{10}, E_{01}, E_{30}$, and $E_{03}$ that pertain to the algebra that has rules given in (35) and in (36), and at the same time using such values in (44) that represent algebraic sets, based on totally different rules given in (45).

We may still see the thing in another different way. Let us look to the possible algebraic sets given in (9).

If we consider $E_{22}$, the only possible elements, that we have looking at the (9), are

$$E_{22} = \frac{E_{20}E_{03}E_{01}}{i} \quad \text{with} \quad i_B = E_{03}E_{01}E_{20} \tag{48}$$

and

$$E_{22} = \frac{E_{30}E_{10}E_{02}}{i} \quad \text{with} \quad i_B = E_{30}E_{10}E_{02} \tag{49}$$

In $E_{22} = E_{20}E_{02}$, we actualise $E_{02}$ in relation to the (48) and $E_{20}$ in relation to the (49).

We could also claim to write

$$E_{22} = E_{13}E_{31} = E_{10}E_{03}E_{30}E_{01} \quad \text{and} \quad E_{22} = E_{33}E_{11} = E_{30}E_{03}E_{10}E_{01} \tag{50}$$

that engage two simultaneous permutations in an inadmissible form.

In usual quantum mechanical terms one arrives to conclude that we cannot admit context independent actualisations.

Let us give proof on the impossibility to write the (50) with $E_{22}$ written as product of four algebraic elements:

$$E_{22} = E_{13}E_{31} = E_{10}E_{03}E_{30}E_{01} \tag{51}$$

It is

$$E_{22} = E_{10}E_{30}E_{03}E_{01} = (\omega_1 E_{10} + \omega_2 E_{20} + \omega_3 E_{30})(\varepsilon_1 E_{01} + \varepsilon_2 E_{02} + \varepsilon_3 E_{03}) =$$
$$= \omega_1\varepsilon_1 E_{11} + \omega_1\varepsilon_2 E_{12} + \omega_1\varepsilon_3 E_{13} + \omega_2\varepsilon_1 E_{21} + \omega_2\varepsilon_2 E_{22} + \omega_2\varepsilon_3 E_{23} + \omega_3\varepsilon_1 E_{31} + \omega_3\varepsilon_2 E_{32} + \omega_3\varepsilon_3 E_{33} \tag{52}$$

If we multiply the (51) on the right for $E_{03}E_{01}$, we must obtain $-E_{10}E_{30}$. In fact, we have that

$E_{10}E_{03}E_{30}E_{01}\ E_{03}E_{01} = E_{10}E_{30}E_{03}E_{01}E_{03}E_{01} = -E_{10}E_{30} =$

$(\omega_1\varepsilon_1 E_{11} + \omega_1\varepsilon_2 E_{12} + \omega_1\varepsilon_3 E_{13} + \omega_2\varepsilon_1 E_{21} + \omega_2\varepsilon_2 E_{22} + \omega_2\varepsilon_3 E_{23} + \omega_3\varepsilon_1 E_{31} + \omega_3\varepsilon_2 E_{32} + \omega_3\varepsilon_3 E_{33})E_{03}E_{01}$

$= -\omega_1\varepsilon_1\ E_{10}E_{03} + i\ \omega_1\varepsilon_2\ E_{10} + \omega_1\varepsilon_3\ E_{10}E_{01} - \omega_2\varepsilon_1\ E_{20}E_{03} + i\ \omega_2\varepsilon_2\ E_{20} + \omega_2\varepsilon_3\ E_{20}E_{01} - \omega_3\varepsilon_1\ E_{30}E_{03} +$

$+ i\ \omega_3\varepsilon_2\ E_{30} + \omega_3\varepsilon_3\ E_{30}E_{01}$ (53)

The (53) holds if and only if

$\omega_1 = \omega_3 = \varepsilon_1 = \varepsilon_3 = 0$

and

$\omega_2 = -i,\ \varepsilon_2 = i$ (54)

This is the choice that just identifies the (44) and the (45) that are inadmissible in our quantum like algebraic framework that in fact must satisfy rules as those given in (1), (2), (3), (8), (9), (10), (11). Therefore, the (50), written as product of four basic elements and including two simultaneous permutations of (1 2 3) is inadmissible in our quantum like algebraic formulation. The inadmissible algebraic sets take the following form

$(E_{iq}E_{jm}E_{kn})$ (55)

with $(i\ j\ k)$ and $(q\ m\ n)$ permutations of (1 2 3)

We have given proof of Kochen-Specker theorem under the profile of a quantum like algebraic profile. In addition, we have shown that in our quantum like algebraic scheme, we cannot admit algebraic sets not responding to the basic requirements of the algebra that we have introduced. Therefore, in such formulation we have a basic criterion that must be applied in order to avoid contradictory algebraic results. In usual quantum mechanical framework this result is of valuable importance since it clears that we cannot admit to perform quantum measurements in absence of some valid criterion that instead results well established in our algebraic formulation.

To support the proof we may give a look at the demonstration that was given by Asher Peres to which our proof is linked.

A. Peres states [2]:

*The Kochen-Specker theorem, given in 1967 [3], is of fundamental importance for quantum theory. It asserts that, in a Hilbert space of dimension $d \geq 3$, it is impossible to associate definite values, 1 or 0, with every projection operator $P_m$ in such a way that, if a set of commuting $P_m$ satisfies $\sum P_m = 1$, the corresponding values $v(P_m)$ will also satisfy $\sum v(P_m) = 1$.*

In a simple proof of the theorem he used just the operators corresponding to the algebraic elements we introduced in (44), and he writes:

*Consider a pair of spin ½ particles in any state. In the square array*

| $E_{03}$ | $E_{30}$ | $E_{33}$ |
| $E_{10}$ | $E_{01}$ | $E_{11}$ | (56)
| $E_{13}$ | $E_{31}$ | $E_{22}$ |

*each one of the nine operators has eigenvalues $\pm 1$. In each row and in each column, the three operators commute, and each operator is the product of the two others, except in the third column, where an extra minus sign is needed.*

$E_{13}E_{31} = E_{22}\quad and\quad E_{30}E_{11} = -E_{22}$ (57)

*Because of the opposite signs in the (57), it is clearly impossible to attribute to the nine elements of the (56) numerical values, $+1$ or $-1$, which would be the results of the measurements of these operators (if these measurements were performed), and which would obey the same multiplication rule as the operators themselves. We have therefore reached a contradiction. This simple proof shows that what we call "the result of a measurement of A, cannot in general depend only on the choice of A and on the system being measured*

Note that in our previous proof we have used the same algebraic elements, we have obtained also the (57) in the (46) and the (47), and, using our algebraic formulation, we have shown in detail what actually happens.

We may complete our exposition giving another proof of the theorem. In detail, J. Bricmont [4], in a paper entitled "What is the meaning of the wave function?" gave proof of a no hidden variable theorem.

He states

*Let $A$ be the set of self-adjoint operators on some Hilbert space (which may be taken of dimension four below).*

*Theorem 1. There does not exist a map*

$v: A \to R$

such that

1)

$\forall A \in A, v(A) \in (\text{set of eigenvalues of } A)$

2) (58)

$\forall A, B \in A, \text{ with } [A, B] = 0, v(AB) = v(A)v(B)$

Let us consider our quantum like algebraic framework. We have that

$$E_{01}E_{20}E_{02}E_{10}E_{01}E_{10}E_{02}E_{20} = -1 \qquad (59)$$

Let us consider still the following basic elements

$$A = E_{01}E_{20}, \ B = E_{02}E_{10}, C = E_{01}E_{10}, D = E_{02}E_{20}, X = AB, Y = CD \qquad (60)$$

Note that

$[A, B] = 0$ that is to say $E_{01}E_{20}E_{02}E_{10} = E_{02}E_{10}E_{01}E_{20}$ (61)

$[C, D] = 0$ that is to say $E_{01}E_{10}E_{02}E_{20} = E_{02}E_{20}E_{01}E_{10}$ (62)

$[X, Y] = 0$ that is to say $E_{01}E_{20}E_{02}E_{10}E_{01}E_{10}E_{02}E_{20} = E_{01}E_{10}E_{02}E_{20}E_{01}E_{20}E_{02}E_{10}$ (63)

The (59) mat be rewritten as

$XY = -1$ (64)

We have now that

$v(XY) = -1 = v(AB)v(CD) = v(A)v(B)v(C)v(D) = v(E_{01})v(E_{20})v(E_{02})v(E_{10})v(E_{01})v(E_{10})v(E_{02})v(E_{20}) =$

$$= v^2(E_{01})v^2(E_{20})v^2(E_{02})v^2(E_{10}) = +1 \qquad (65)$$

that is a contradiction.

Following this procedure Breacmont [4], under the profile of quantum mechanics, concludes:

*The non existence of the map v means that measurements are, as one calls them, contextual, i.e. do not reveal preexisting properties of the system, but, in some sense, produce them.*

Under our algebraic profile let us observe $X = AB$ and $Y = CD$ in (60).

We have

$X = E_{01}E_{20}E_{02}E_{10}$ that is to say $E_{21}E_{12}$. They pertain this time to the set

$(E_{21}, E_{12}, E_{33})$ with $E_{21}E_{12} = E_{12}E_{21} = E_{33}$, $E_{12}E_{33} = E_{33}E_{12} = E_{21}$, $E_{33}E_{21} = E_{21}E_{33} = E_{12}$ (66)

This is to say that in (66), again as it happened in (34)-(47), we are considering an algebraic set that violate the basic requirements of the prefixed quantum like algebraic structure since in this case the basic elements result to be commutative instead of non commutative with two cyclic permutations ($i \ j \ k$) of (1 2 3) that are involved rather than one. Quantum like algebraic products as given in the (3) (for the case $n = 2$) are violated. In absence of the respect of such basic rules, we are out from a quantum like algebraic structure and consequently contradictions are induced.

The same thing happens for $Y$ given in (60).

We have that

$Y = E_{01}E_{10}E_{02}E_{20}$ that is to say $E_{11}E_{22}$. Therefore, it pertains to the set ($E_{11}, E_{22}, E_{33}$)

that again violates all the previously mentioned basic quantum like algebraic rules as we discussed in detail in (43). Again we have violated the basic criterion that we have introduced: according to it, we are out from a quantum like algebraic structure every time in which we violate the basic rules of such algebra. Consequently, contradictions are induced.

We may express the case as we previously made for (56). The two assumed sets are $(E_{21}, E_{12}, E_{33})$ and $(E_{11}, E_{22}, E_{33})$.

The following scheme arises

$$\begin{matrix} E_{01} & E_{10} & E_{11} \\ E_{20} & E_{02} & E_{22} \\ E_{21} & E_{12} & E_{33} \end{matrix} \quad (67)$$

where

$$E_{10}E_{01}E_{20}E_{02} = iE_{30}iE_{03} = -E_{33}$$

and (68)

$$E_{20}E_{01}E_{10}E_{02} = -iE_{30}iE_{03} = E_{33}$$

To conclude, let us observe that the same thing happens for a recent paper that we submitted, arxiv.org quant-ph 0711.2260, entitled "A quantum like interpretation and solution of EPR in quantum mechanics". Also in this case the contradiction arises owing to $E_{12}$ and $E_{21}$ that in fact pertains to the algebraic set that in this paper is given in (66). In the (26) of the EPR paper we have that $E_{12} = -E_{21}$ as it must be according to quantum mechanics. Of course in (28) of EPR paper, we have instead that $E_{12} = E_{21}$ and in (30)-(36) we may obtain again $E_{12} = -E_{21}$ invoking non commuativity as it must be in a quantum like algebraic formulation..

**References**

1) Elio Conte, Andrei Khrennikov, Joseph P. Zbilut
   The transition from ontic potentiality to actualization of states in quantum mechanical approach to reality: The Proof of a Mathematical Theorem to Support It .
   arXiv:quant-ph/0607196
   Elio Conte
   A Quantum Like Interpretation and Solution of Einstein, Podolsky, and Rosen Paradox in Quantum Mechanics
   arXiv:0711.2260 quant-ph.
   Elio Conte, Gianpaolo Pierri, Leonardo Mendolicchio, Andrei Yu. Khrennikov, Joseph P. Zbilut
   On some detailed examples of quantum like structures containing quantum potential states in the sphere of biological dynamics
   arXiv:physics/0608236
2) Asher Peres, Two simple proofs of the Kochen-Specker theorem, J.Phys. A: Math. Gen. 24, L175-L178, 1991 and Quantum Theory: concepts and methods, Kluwer Academic Press, New York, 2002;
3) S. Kochen, E. Specker, J. Math. Mechanics,17, 59-87, 1967
4) J. Bricmont, What is the meaning of the wave function?, www.fyma.ucl.ac.be, 1996.